\documentclass[twocolumn,amsmath,amssymb,aps,prl,reprint,groupedaddress]{revtex4-1}
\usepackage{graphicx}
\usepackage{float}
\usepackage[caption=false]{subfig}
\usepackage{tensor}
\newcommand{\mathsym}[1]{{}}
\newcommand{\unicode}[1]{{}}
\usepackage[usenames]{color}
\usepackage[subpreambles=true]{standalone}

\begin{document}

\title{Ground States of Crystalline Caps: Generalized Jellium on Curved Space}


\author{Siyu Li}

\author{Roya Zandi}
\affiliation{Department of Physics and Astronomy,
   University of California, Riverside, CA 92521, USA}
\author{Alex Travesset}

\affiliation{Department of Physics and Astronomy, Iowa State University and Ames Lab, Ames, IA 50011, USA}

\author{Gregory M. Grason}
\affiliation{Department of Polymer Science and Engineering, University of Massachusetts, Amherst, MA 01003, USA}
\begin{abstract}
We study the structure and elastic energy of the ground states of crystalline caps conforming to a spherical surface. These ground states consist of positive disclination defects in structures spanning from flat and weakly curved crystals to closed shells. We compare two different continuum theories and one discrete-lattice approach for the elastic energy of defective caps.   We first investigate the transition between defect-free caps to single-disclination ground states and show that it is characterized by continuous symmetry-breaking transition.  Further, we show that ground states with icosahedral subgroup symmetries in caps arise across a range of curvatures, even far from the closure point of complete shells.  Thus, while superficially similar to other models of 2D ``jellium'' (e.g. superconducting disks and 2D Wigner crystals), the interplay between the free edge of crystalline caps and the non-Euclidean geometry of its embedding leads to non trivial ground state behaviors, without counterpart in planar jellium models. 

\end{abstract}
\date{\today}

\maketitle



Spherical crystals are elementary models of geometric frustration in materials, with important realizations in a range of systems from fullerenes and protein shells to particle- and molecular-coated droplets~\cite{caspar,ful,bruinsma_pnas,Sanaz2018,bausch, irvine,sloutskin}.  The long-standing problem of finding the ground state of $N$ particles covering the sphere, known as the generalized Thompson problem~\cite{smale,moore}, derives its complexity from the basic conflict between equi-triangular order and positive Gaussian curvature, that is, elliptic geometries~\cite{nelson_book,kleman,bowick,lucks}. For closed shells, topology dictates the total charge of disclinations (i.e. sites deviating from six-fold coordination) to $\sum_{i=1}^{n_d} q_i = 12 $, which for the simplest case of only 5-fold defects ($q_i=+1$) constrains the number of disclinations to $n_d=12$~\cite{Wagner2015,bowick,giomi_bowick}. Considerable progress has been made by numerically optimizing, classifying and rationalizing the patterns of defects in an otherwise six-fold, quasi-triangular order of closed shells~\cite{altschuler,moore,bowick_caccuito,wales_09}. In contrast, the structure and energetics of defect ground states of partially-closed crystalline shells, or {\it crystalline caps}, which span the gap between defect-free planar crystals and closed shells, remain unresolved~\cite{azadi_prl, azadi_pre, Castelnovo2017, li}. 


\begin{figure}[b]
\centering
\includegraphics[width=\linewidth]{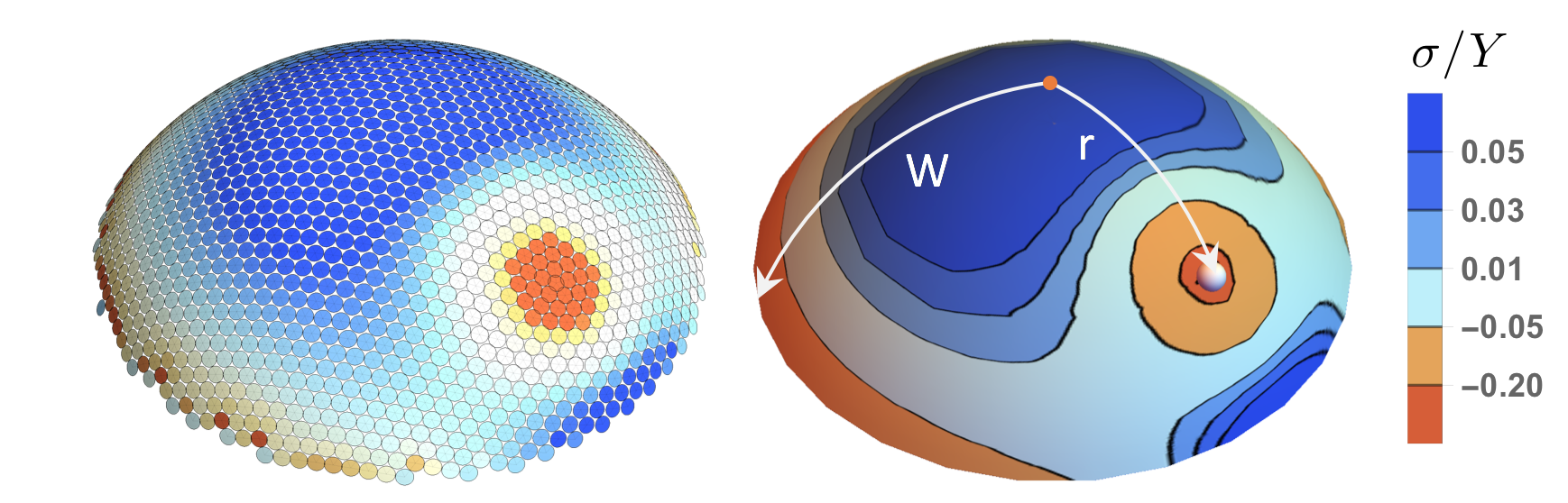}
\caption{Map stresses in an crystalline cap with an off-center 5-fold disclination located at $r=0.6W$ from  see Eq.~\ref{Eq:Df_elastic_discrete} (left) and covariant theory, see Eq.~\ref{Eq:bieq} (right). }
\label{stress}
\end{figure}

Unlike closed shells, the number of defects in the interior of crystalline caps is not topologically constrained. Disclinations can be created and destroyed in the free boundary of the cap, adjusting their number in accordance with energetic considerations deriving from favorable elastic screening of curvature-induced stresses~\cite{schneider,giomi,irvine,grason_prl_10,grason_pre_12}. In this vein, ground states of crystalline caps may be described by a {\it generalized jellium} (GJ) model, in which both Gaussian curvature and disclinations act as  point-like sources of ``elastic stress" mediating their energetics~\cite{nelson_book}.  While the most familiar examples of GJ describe bulk phases like the Wigner crystal~\cite{maradudin, fisher} or Abrikosov lattice of type-II superconductors~\cite{geim,baelus,schweigert} , many physical scenarios are described by finite domains of homogeneous ``charged" backgrounds punctuated by a small number of neutralizing ``point charges". For example, under an increasing magnetic field, the ground state wave function of 2D superconducting discs exhibit a complex sequence of continuous and discontinuous transitions in the number and symmetry of vortices~\cite{baelus,schweigert}, which derives from the generic incommensurability of the net applied flux with the quantized flux per elementary vortex.  

Although superficially similar to planar GJ models, crystalline caps are distinguished by their non-Euclidean (elliptic) geometry which alters the shape and length of the boundary free cap. Because free boundaries have the ability to screen defect-induced stress, the non-planar geometry of caps has a critical impact on the ground state order of its defects. In this paper, we describe the spectrum of defect ground states of crystalline caps using a combination of continuum elasticity theory and simulation models.  We show that geometrically nonlinear effects at the free edges give rise to a novel continuous transition from the defect-free interiors to a trapped central disclination as well as the existence of soft, near-edge trapping of low-energy defects.  This latter mechanism leads to a non-trivial sequence of defect ground states, which is characterized by a non-monotonic dependence of number of defects in the cap interior on the integrated Gaussian curvature, or sphere area coverage. Finally, we show that the non-Euclidean geometry of the cap edge is essential for the emergence of defect ground states with symmetries commensurate with the icosahedral symmetry of closed shells (e.g. the Thompson problem~\cite{bowick_caccuito}).

To obtain the elastic energy ground states of a cap, we consider a 2D hexagonal crystal on a sphere of radius $R$ extending up to a radial distance $W=\theta_c R$ from its center to its free edge, see Fig.~\ref{stress} and S6.  While brittle crystallization on spherical surfaces is known to lead to elastic instabilities of the free boundary shape~\cite{schneider,bruinsma,manoharan,hall_intfocus_17}, here we consider the limit of large edge energy~\footnote{We expect this boundary shape instability is suppressed in the limit of edge energies much larger than $\sim Y\theta_c^4 W$~\cite{bruinsma,hall_intfocus_17,Reiss_nucleation}.}, where the boundary of the crystal remains axisymmetric~\cite{Reiss_nucleation,li,Sanaz2018,Wagner2015}. Furthermore, we restrict our attention to ground states that possess only $q_i=+1$ (five-fold) disclination defects in an otherwise hexagonal bulk order, neglecting the possibilities of dislocations chains or ``scars" that become energetically favorable when the lattice spacing $a$ (and dislocation core energy) becomes sufficiently small compared to $W$~\cite{bowick,azadi_prl}.

The elastic energy takes the form
\begin{equation}\label{F1}
F=\frac{1}{2 Y}\int d^2{\bf x} \sqrt{g}\left(\Delta\chi\right)^2,
\end{equation}
where $\sqrt{g}$ is the metric induced in the spherical cap and $\chi$ is the Airy stress function, which encodes the elastic intra-crystal stress (see SI Eqs. (S9-S11)), and satisfies~\cite{bowick}
\begin{eqnarray}\label{Eq:bieq}
\frac{1}{Y}\Delta ^2\chi({\bf x}) &=&  s({\bf x})- K({\bf x}),
\end{eqnarray}
where $K({\bf x})$ is the Gaussian curvature, $\Delta = \frac{1}{\sqrt{g}}\partial_{i}(\sqrt{g}g^{ij}\partial_{j})$ is the Laplacian on a 2D surface with metric $g_{ij}$ and 
\begin{equation}
s({\bf x}) =\frac{\pi}{3\sqrt{g}} \sum_{\alpha=1}^{n_d} q_\alpha \delta({\bf x}- {\bf x}_\alpha) 
\end{equation}
is the disclination density, composed of $n_d$ disclinations possessing $q_\alpha =+1$, the topological charge per single 5-fold defect (i.e. with wedge angle  $\pi q_\alpha/3$ per defect).  The crucial effect of the free boundary motion is captured by imposing vanishing normal stress at the cap edge on solutions for $\chi$.  While strictly speaking topology requires a fixed balance between disclinations in the bulk and on the boundary of the cap~\cite{giomi_bowick}, stress screening by free boundary implies that the elastic effect of defects as they approach the boundary, hence, energetics are sensitive only to {\it interior defects}.

The elastic energy of multi-disclination configurations in caps were previously computed~\footnote{The derivations for FvK in axisymmetric 2D crystals were originally applied to twisted columnar phases, the elastic energy of which is identical to crystalline caps in the small-slope, or low-twist, regimes.} for the so-called {\it F\"oppl v\'an K\'arman} (FvK) limit~\cite{grason_prl_10,grason_pre_12}, strictly justified in the limit of small-slopes~\cite{landau}.  The FvK limit corresponds to approximating the metric $g_{ij}$ to be planar in Laplacian while retaining $K({\bf x})=1/R^2$ as a homogeneous source for Airy stress on the right hand side of Eq.~(\ref{Eq:bieq}). While the FvK theory is tractable for axisymmetric caps with arbitrary defect arrangements~\cite{grason_jcp_16,Castelnovo2017}, the small-slope limit is not satisfied for most of curvatures where defects are energetically favored.  Recently, a rigorous framework for the fully covariant elasticity theory caps has been developed~\cite{li, Li2019}, which builds from elements in the theory of incompatible elasticity~\cite{Efrati2009,MosheSharonKupferman2015}, and more crucially, allows for the computation of the elastic energy of multi-defect configurations with arbitrary complexity (See SI). This covariant approach, which hereon we will refer as LF (Lagrange Formalism)~\cite{Li2019}, captures the full geometric non-linearity of the cap shape through incorporation of the spherical metric in the deformed state (i.e. in so-called azimuthal-equidistant coordinates $g_{rr} = 1, g_{r\phi} = 0 , g_{\phi \phi} = R^2 \sin^2(r/R) $) while evaluating the Laplacian  and area integral in Eq.~(\ref{F1}).

In the context of the GJ models, the FvK model may be considered as the biharmonic analogy to the 2D ``electrostatic'' theory of superconducting disks, that is, generalized by the much longer range interactions between monopoles in the biharmonic theory (i.e. monopole interactions with separation $r$ grow as $\sim r^2\ln r$, in comparison to $\ln r$ for 2D Coulomb~\cite{romanov}).  In comparison, the fully covariant theory embeds the ``biharmonic electrostatics'' problem in a non-Euclidean spherical geometry.  The effect of the reduced geodesic separation between disclinations embedded on spheres, in combination with relative reduction of the perimeter to domain size ratio for caps compared to disks with the same radial distance-- both captured in the covariant theory -- qualitatively alters the structure and energetics of disclinations in caps, as we will show below.

The accuracy of the different approximations to the continuum elastic theory can be tested in comparison to simulations of the lattice model introduced by Nelson and Seung (NS)~\cite{nelson_seung} 
\begin{equation}\label{Eq:Df_elastic_discrete}
E_{\rm NS} =\frac{k}{2} \sum_{\langle i, j \rangle} \left(| {\bf r}_i-{\bf r}_j|-a\right)^2 \ ,
\end{equation}
which consists of a triangular network of nearest neighbor springs of rest length $a$ and spring constant $k$. As described in the SI, configurations possessing up to $n_d=0, 1 ,2$ or $3$ five-fold disclinations are constructed by introducing multiple $60^\circ$ Volterra cuts, generalizing previous approaches to disclinated meshes~\cite{Travesset2003,grason_prl_10}. Figure~\ref{stress} illustrates that there is an excellent agreement between the stress distribution computed from the LF and simulation of a cap with a single off-center disclination, see SI for the details.  


We first describe the elementary transition from the stable defect-free ground state to the ground state possessing the first stable internal disclination, with  ($q=+1$).  For the FvK theory the elastic energy (per unit area) of the single-disclination cap is known, and is a function of aperture angle $\theta_c = W/R$, off-center defect position $r$ and disclination charge $q$~\cite{grason_pre_12}
\begin{equation}
\label{eq: EFvK}
    \frac{E_{\rm FvK}}{Y A} = \frac{\theta_c^4}{384}+ \frac{1}{96}\Big(\frac{q^2}{3} - \frac{q \theta_c^2}{2} \Big) \Big[1-(r/W)^2 \Big]^2 ,
\end{equation}
where the first term derives from curvature-generated stress, the second, from the elastic self-energy of the disclination and the third, from the mutual elastic screening of the curvature and disclinations stress~\footnote{Note that we neglect in this analysis the core energy of disclinations, which is expected on dimensional grounds to be comparable to $~Ya^2$ and hence much smaller than the far-field elastic energies of disclinations, $\sim YW^2$}. The analytic formula for the elastic energy for the covariant energy, $E_{\rm LF}$, is more complex as shown the in the SI and described in ref.~\cite{Li2019}, but it reduces to eq. (\ref{eq: EFvK}) in the limit of small curvature, $\theta_c \ll 1$.

\begin{figure}
 \centering
 \includegraphics[width=\linewidth]{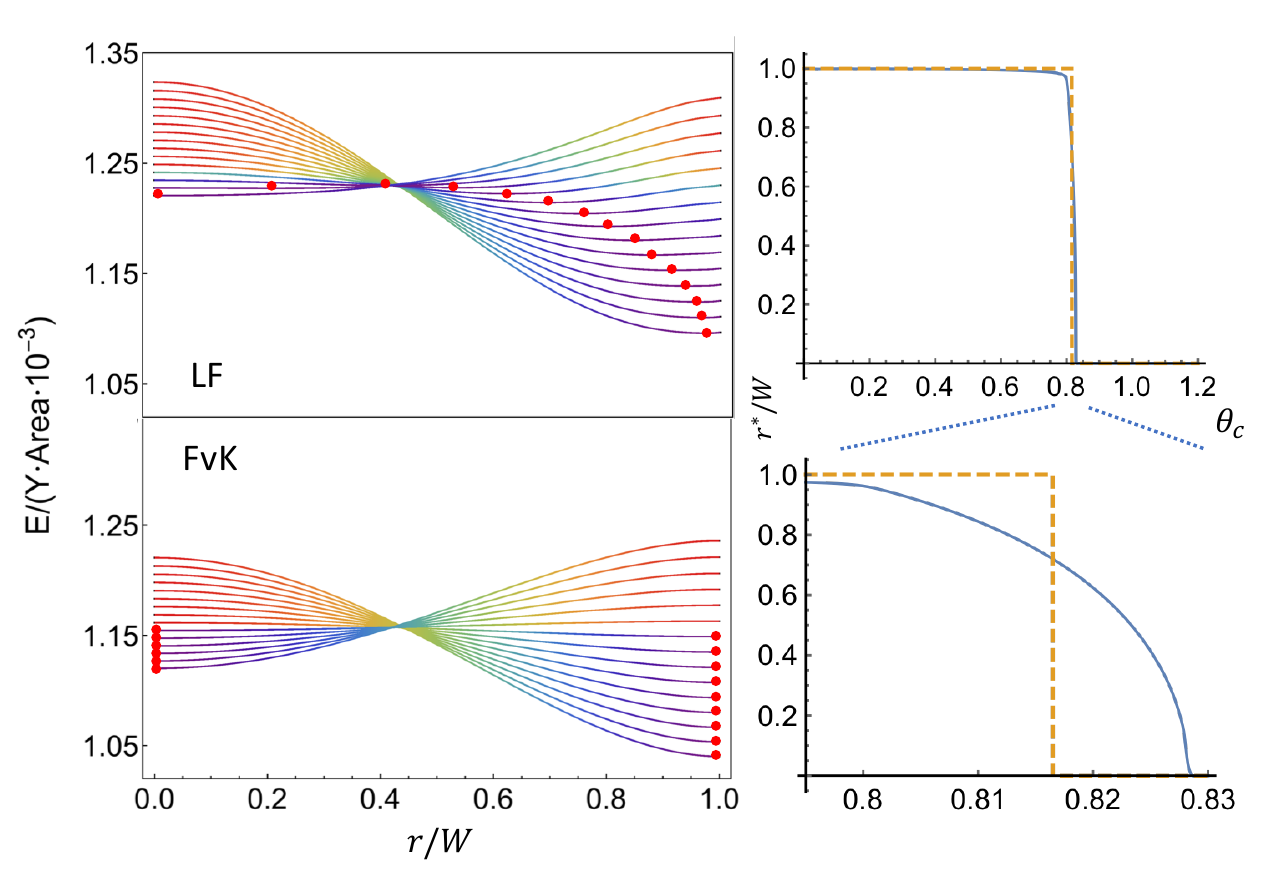}
  \caption{\footnotesize Energy density vs. the location of a disclination for different cap sizes.  The red dots indicate the location of disclination that minimizes the energy density.  Based on FvK, there is a first order transition from the edge to the center between $\theta=0. 795$ and $\theta=0.83$.  However, there is a second order (smooth) transition from the edge to the center according to covariant theory. }
\label{transition}
\end{figure}

Because the self-energy and defect-curvature interaction contributions exhibit exactly the same dependence on defect position $r^*$ in the FvK theory, this theory predicts a simple 1st-order transition from the defect-expelled state (minimum at $r^*=W$) for $\theta_c < \theta^*$ to the defect-centered state (minimum at $r^*=0$) for $\theta_c > \theta^*$ at a critical cap angle $\theta^*=\sqrt{2/3}\approx 0.816$, see Fig.~\ref{transition}. Note that a standard heuristic argument~\cite{irvine} that considers the cap angle, $\theta_n$, at which integrated Gaussian curvature ``neutralizes'' single-five fold defects yields
\begin{equation}\label{Eq:topology}
\int dA ~s({\bf x})=\int dA~ K({\bf x}) \rightarrow \frac{\pi}{3}=2\pi(1-\cos\theta_n) \ ,
\end{equation}
or $\theta_n=\arccos(5/6)\simeq 0.59$. This value falls well below the above prediction for $\theta^*$, revealing that the free cap boundary requires significant ``overcharging'' of curvature-induced stress to overcome the elastic self-energy of the monopole disclination charge.  

Figure~\ref{transition} illustrates the elastic energy  vs. defect position predicted by the covariant theory, where it is found that the first disclination is expelled continuously from the boundary, starting approximately at $\theta \simeq 0.795$ reaching the center of the disk ($r^*=0$) at around $\theta \simeq 0.83$, exhibiting a range of off-center defect equilibria $0<r^*<W$ in this narrow curvature window\footnote{We point out that in Ref.~\cite{li}, the appearance of the first defect was found to be at $\theta \simeq 0.66$ instead of $0.795$ due to numerical error arising from summing over not enough modes in the multipole expansion of self-energy. A discussion is provided in SI.}. 

The distinct first- vs. second-order nature of transitions respectively predicted by FvK and covariant theory highlights the impact of the non-Euclidean geometry of the cap. While the FvK theory predicts the self-energy and defect-curvature interactions to exactly balance at the critical curvature, the appearance of stable off-center equilibrium for disclinations in the covariant theory can be associated with the imbalance between these two competing effects.  Figure~S2 plots the relative magnitudes of these terms for the covariant theory and FvK, showing that the defect self-energy is relatively depressed, while the defect-curvature is enhanced, as disclinations approach the free edge of the cap in the covariant theory. The enhanced curvature-generated attraction at the edge can be associated from the geometrically nonlinear edge compression, which grows faster than quadratic approximation of the small-slope FvK theory.  On the other hand, the near-edge weakening of the self-energy that tends to expel defects from the cap can be associated with the smaller geodesic curvature of the cap edge compared to a planar disk, which amplifies the far-field screening of defect stresses by free boundary relaxation (see SI Sec. I.C for discussion).  These two effects of the non-Euclidean geometry of caps conspire to create ``soft traps'', stabilizing off-center equilibrium.  

We note that the simulation model (Fig.~S4c) shows a transition from defect-free to centered-disclination ground state at a cap angle quite close to both continuum calculations, $\theta^* \simeq 0.84$. However, discreteness effects associated with the finite core size (zone of anharmonic strain) and non-circularity of the free edge obscure resolution on the near-edge elastic binding of single defects. 

\begin{figure*}
\centering
\includegraphics[width=1\linewidth]{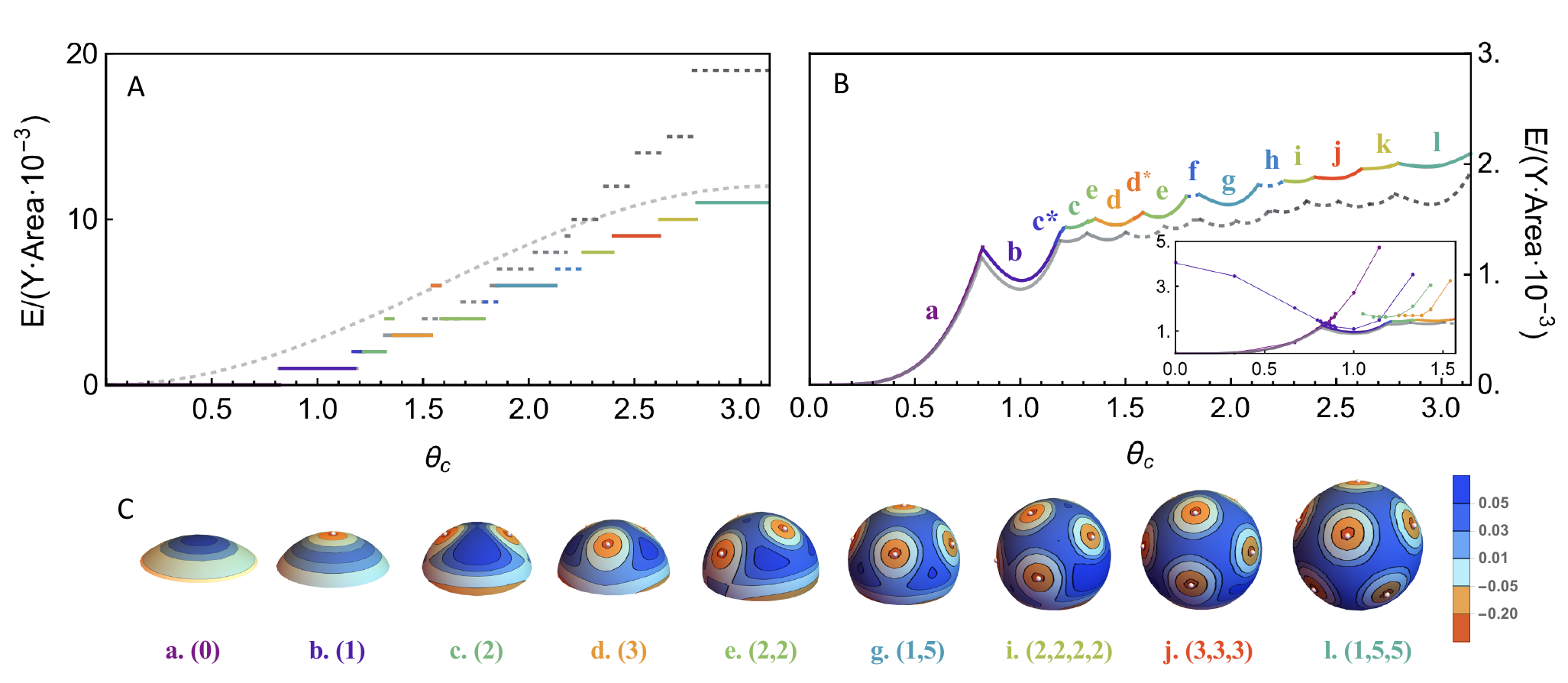}
\caption{\footnotesize Comparison of ground state energy and defects number in FvK vs. LF. The colorful and dark gray lines correspond to the ground states obtained from LF and FvK theories. Only solid lines indicate the defect configurations follow IO symmetry.
(a) Number of defects as a function of the growing cap size $\theta_c$. The dot-dashed line denotes to the topological charge neutrality condition.
(b) Ground state energy as a function of cap size.  The inset compares the ground state energy in simulation (dots) to LF (solid lines) and FvK (gray lines). 
(c) Stress distribution of the icosahedral subgroups emergent as the ground states in covariant theory.}
\label{gallery}
\end{figure*}


We now consider the evolution of multi-defect ground state structure with cap growth from nearly-flat ($\theta_c \lesssim 1$) to fully closed shells ($\theta_c \to \pi$).  Numerous previous considerations of the Thompson problem and its variants~\cite{bowick_caccuito, giomi_bowick} point towards {\it icosahedral} defect configurations as ground states in closed spheres.  Yet, it remains unclear whether, and at what point, ground state structures of incomplete caps conform to this symmetry.  Thus, we compare the elastic energy of two basic classes of ground state symmetries: the first is point pattern of disclinations possessing subgroup symmetries of icosahedron (IO), as illustrated in a stereographic projection in Fig.~S8; and second are point patterns composed of concentric $n$-fold symmetric rings of defects with composite symmetries listed in Table~S1. For each configuration, we compute the minimum energy by varying the arc-radial distance of concentric rings defects, keeping the azimuthal separation of defects fixed in each ring. For the second class of configurations, we also minimize with respect of the rotation angle between concentric rings.   For example, a (2,4) configuration is composed of 2 defects in the first ring and 4 defects in the second, in both cases evenly distributed. Such a configuration is then minimized with respect to  the arc radius of two rings $r_1$, $r_2$ and the phase rotation angle $\phi$ between the two rings. For the IO class, while we minimize over arc-radii of rings, we retain the fixed angular correlations between consecutive rings deriving from icosohedral symmetry (e.g. defects on consecutive rings in the 3-fold IO state are separated by 60$^\circ$).

We plot the results in terms of the number of {\it bulk} disclinations and energy density of ground state configurations in Fig.~\ref{gallery} A and B respectively, for both the FvK (black) and LF (color) elastic theories.  Since their elastic energy contributions become arbitrarily small as defects approach the cap boundary, we introduce a cut off radius of $0.95 W$, beyond which we count defects to be at the boundary of the crystal and not in the bulk.  

The ground state energies of the LF increases monotonically, with few exceptions (see $(2,2)$ and $(3,3)$ following $(2)$ and $(3)$, respectively, due to the weakly, near-edge defect traps) and the number of disclinations always remain below the condition of topological charge neutrality (i.e. a number of internal defects whose total defect angle is equal to the integrated Gaussian curvature), with the calculation converging to the topological correct condition Eq.~\ref{Eq:topology} of twelve defects with icosahedron symmetry.  In contrast, for $\theta_c \gtrsim 2$ the ground states of the FvK model begin to greatly exceed the topological condition Eq.~\ref{Eq:topology}, eventually growing to +19 disclinations in the complete shell, far in excess of the +12 required on a closed sphere.  In terms of the energy density (Fig.~\ref{gallery}B), both theories show a similar crossover from the $\sim \theta_c^4$ growth of elastic energy for defect free caps to the plateau-like series of multi-disclination minima at large coverage.  Notwithstanding these qualitative similarities in the curvature-dependence of the energy, the ground states symmetries of the two models differ substantially.  As illustrated in Fig.~\ref{gallery}C, all but two ground states of the LF possess quasi-icosahedral symmetry. In contrast, as shown in SI Fig.~S7, ground states of the FvK theory with $n_d>3$ break icosahedral symmetry, with the sole exception of a narrow range of stable $(3,3)$, exhibiting higher-fold concentric-ring defect patterns that are also characteristic of planar vortex packings in superconducting ground states~\cite{baelus}. 

In summary, the detailed comparison of the widely-used elastic plate theory (FvK), a covariant continuum elasticity theory (LF) and fully nonlinear, discrete lattice model (NS) of crystalline caps reveal that qualitative features of the ground state structure and energetics derive from the non-Euclidean embedding of caps, and crucially, their free boundaries.  Beyond its effects to modify the transition from defect-free and defective caps for relatively small curvature, we find spherical geometry is essential for the emergence of icosahedral defect configurations, even at cap curvatures surprisingly far from closure.  Whereas the FvK model (i.e. ``biharmonic, planar jellium'') predicts defect numbers and arrangements wildly departing from structures predicted in closed shells (i.e. $n_d(\theta_c \to \pi) \gg 12$), the covarient LF model (i.e. "biharmonic, spherical jellium'') yields ground states that smoothly interpolate to the grounds of twelve $q=+1$ defects for $\theta_c \to \pi$.  This result demonstrates that the topological constraint that emerges for {\it closed shells} (i.e. $\sum_{i=1}^{n_d} q_i = 12$) is not strictly necessary for the emergence of icosahedral defect arrangements, and such ground states can emerge purely from the (topologically-unconstrained) energetics of multi-defect arrangements on {\it incomplete shells}, but only when embedded properly in a spherical geometry.

The authors are grateful to the hospitality of Aspen Center for Physics (NSF PHY 1607611) where this work began.  The work of SL and RZ was supported by the NSF through Awards DMR-1719550. AT acknowledges support from DMR-1606336 and GG from the DOE Office of Science, Basic Energy Sciences, under Award No. DE-SC0017870.

\bibliography{bibfile}

\end{document}


\title{ Supplementary Information: \\
Ground States of Crystalline Caps: Generalized Jellium on Curved Space}
\author{Siyu Li$^1$, Roya Zandi$^1$, Alex Travesset$^2$, and Gregory M. Grason$^3$}
\affiliation{$^1$Department of Physics and Astronomy,
   University of California, Riverside, CA 92521\\
$^2$Department of Physics and Astronomy, Iowa State University and Ames Lab, Ames, IA 50011\\
$^3$Department of Polymer Science and Engineering, University of Massachusetts, Amherst, MA 01003}

\date{\today}
\maketitle

\section{Continuum elastic theories for multi-disclination caps}

Here we briefly summarize formulas used to compute the elastic energy of multi-disclination caps in the FvK and the covariant theory (LF). Full details of the derivation can be found in refs.~\cite{grason_pre_12} and ~\cite{Li2019}, respectively.

\subsection{F\"oppl v\'an K\'arman (FvK) energy}

The elastic (in-plane) stress $\sigma_{ij}$ for the FvK energy derives from the standard 2D anti-symmetric derivatives Airy stress $\chi$ (in polar coordinates),

\begin{eqnarray}
\sigma^{rr}_{\text{FvK}}&=&\frac{1}{r}\partial_r\chi+\frac{1}{r^2}\partial_{\phi\phi}\chi \\
\sigma^{r\phi}_{\text{FvK}}&=&-\partial_r(\frac{1}{r}\partial_\phi\chi)\\
\sigma^{\phi\phi}_{\text{FvK}}&=&\partial_{rr}\chi
\end{eqnarray}

Using these relations, Eq. (2) in the main text can be solved for any arbitrary arrangment of disclinations (with topological charge $q_\alpha$ and positions ${\bf x}_\alpha$), subject to vanishing radial stress at the free cap edge, i.e. $\sigma_{rr}=\sigma_{r\phi}=0$ at $r=W$, see Fig. S6.  This results in a total elastic energy per unit area
\begin{align}\label{F1}
E_{\rm FvK}/A &= \frac{Y}{\pi W^2}\int d^2 {\bf x} \big(\Delta \chi\big)^2  \nonumber \\
&=\epsilon^{\rm(FvK)}_{gg}(\theta_c) + \sum_{\alpha=1}^N \epsilon^{\rm(FvK)}_{gd}({\bf x}_\alpha, \theta_c) + \sum_{\alpha=1}^N \epsilon^{\rm(FvK)}_{dd-self}({\bf x}_\alpha) + \sum_{\alpha=1}^N \sum_{\beta>\alpha}^N \epsilon^{\rm(FvK)}_{dd-int}({\bf x}_\alpha, {\bf x}_\beta) ,
\end{align}
where $\Delta \chi=r^{-1} \partial_r(r \partial
_r \chi) + r^{-2} \partial_\phi^2 \chi$ is the 2D planar Laplacian.  The first term derives from the elastic energy density of curving a defect-free cap
\begin{equation}
\epsilon^{\rm(FvK)}_{gg}(\theta_c)=\frac{Y \theta_c^4}{384}
\end{equation}
while the second derives from the ``elastic interaction'' between Gaussian curvature and defect-generated stresses
\begin{equation}
\epsilon^{\rm(FvK)}_{gd}({\bf x}_\alpha, \theta_c)=-\frac{Y q_\alpha \theta_c^2}{192}\Big[1-(r/W)^2 \Big]^2 ,
\end{equation}
and the third describes the elastic ``self'' interaction of disclinations 
\begin{equation}
\epsilon^{\rm(FvK)}_{dd-self}({\bf x}_\alpha, \theta_c)=\frac{Y q_\alpha^2 }{288}\Big[1-(r/W)^2 \Big]^2 .
\end{equation}
The final term describes pairwise, elastic interactions between disclinations in caps,
\begin{equation}
\epsilon^{\rm(FvK)}_{dd-int}({\bf x}_\alpha, {\bf x}_\beta)= \frac{Y q_\alpha q_\beta}{32}\Bigg\{\Big(1-\frac{r_\alpha^2}{W^2}\Big) \Big(1-\frac{r_\beta^2}{W^2}\Big) + \frac{|{\bf x}_\alpha-{\bf x}_\beta|^2}{W^2} \ln \bigg[ \frac{|{\bf x}_\alpha-{\bf x}_\beta|^2}{(W^2-r_\alpha^2)(W^2-r_\beta^2)/W^2+|{\bf x}_\alpha-{\bf x}_\beta|^2 }. \bigg] \Bigg\}
\end{equation}

\subsection{Covariant Energy: LF model}

In the covariant theory, the relationship between the Airy stress $\chi$ and the stress-tensor $\sigma_{ij}$ in the crystal has the form,
\begin{eqnarray}
  \sigma^{rr}_{\text{LF}} &=& \frac{1}{R} \cot(r/R) \partial_r \chi + \frac{1}{R^2 \sin^2(r/R)}\partial_{\phi\phi} \chi \\
  \sigma^{r\phi}_{\text{LF}} &=& -\frac{1}{R\sin(r/R)} \partial_r \left( \frac{1}{R\sin(r/R)} \partial_{\phi} \chi \right) \\
  \sigma^{\phi\phi}_{\text{LF}} &=& \frac{1}{R^2\sin^2(r/R)} \partial_{rr} \chi,
\end{eqnarray}
where differences from the FvK formulation derive from covariant derivatives on the spherical metric~\cite{Li2019}.

The energy in LF can be written as
\begin{align}\label{F2}
E_{\rm LF}/A &= \frac{Y}{4\pi(1-\cos \theta_c)}\int d^2{\bf x} d^2{\bf y} \left[(K({\bf x})-s({\bf x}) )G({\bf x},{\bf y}) (K({\bf y})-s({\bf y})) \right] \nonumber\\
&=\epsilon^{\rm(LF)}_{gg}(\theta_c) + \sum_{\alpha=1}^N \epsilon^{\rm(LF)}_{gd}({\bf x}_\alpha, \theta_c) + \sum_{\alpha=1}^N \epsilon^{\rm(LF)}_{dd-self}({\bf x}_\alpha) + \sum_{\alpha=1}^N \sum_{\beta>\alpha}^N \epsilon^{\rm(LF)}_{dd-int}({\bf x}_\alpha, {\bf x}_\beta) .
\end{align}
Here the curvature induced elastic energy takes the form,
\begin{equation}
\epsilon^{\rm(LF)}_{gg}(\theta_c)=\frac{Y}{4\pi(1-\cos \theta_c)}\int d^2 {\bf x} ~\Omega({\bf x}) , 
\end{equation}
while the defect-curvature interaction has the following form,
\begin{equation}
\epsilon^{\rm(LF)}_{gd}({\bf x}_\alpha, \theta_c) = -\frac{Y}{6(1-\cos \theta_c)} q_\alpha ~\Omega({\bf x}_\alpha),
\end{equation}
with,
\begin{align} \label{eq:omega}
\Omega(r)=&\text{Li}_2(-\tan(\frac{r}{2})^2)+\frac{1}{2} \log(1+\tan^2(\frac{r}{2}))^2 +\log(1+\tan^2(\frac{r}{2}))
\cot^2(\frac{\theta_c}{2}) \log(1+\tan^2(\frac{\theta_c}{2})) \nonumber\\
&-\text{Li}_2(-\tan(\frac{\theta_c}{2})^2)-\frac{1}{2} \log(1+\tan^2(\frac{\theta_c}{2}))^2 -\log(1+\tan^2(\frac{\theta_c}{2}))
\cot^2(\frac{\theta_c}{2}) \log(1+\tan^2(\frac{\theta_c}{2})) .
\end{align}
Further, the disclination self-energy,
\begin{equation}
\epsilon^{\rm(LF)}_{dd-self}({\bf x}_\alpha) = \frac{Y}{72(1-\cos \theta_c)}  q_\alpha^2 G({\bf x}_\alpha,{\bf x}_\alpha) ,
\end{equation}
and the elastic interactions between disclinations,
\begin{equation}
\epsilon^{\rm(LF)}_{dd-int}({\bf x}_\alpha, {\bf x}_\beta) = \frac{Y}{36 (1-\cos \theta_c)} q_\alpha q_\beta G({\bf x}_\alpha,{\bf x}_\beta),
\end{equation}
are both obtained from the Green's function,
\begin{align}\label{G2}
G({\bf x},{\bf x}_i)=&\text{Li}_2 (-\tan^2(\frac{r_>}{2})) +\log(\tan(\frac{r_>}{2}))\left(\log(1+\tan^2(\frac{r_>}{2}))+\log(1+\tan^2(\frac{r_<}{2})\right) \nonumber\\
&+\sum _{n=1}^{\infty}\frac{1}{n}\left(\frac{1}{n}\frac{\tan^n(\frac{r_<}{2})}{\tan^n(\frac{r_>}{2})} -H_n(\cot(\frac{r_>}{2}))\tan^n(\frac{r_<}{2}) -\cot^n(\frac{r_>}{2}) H_n(\tan(\frac{r_<}{2}))\right)
\cos(n\phi) \nonumber\\
&+U({\bf x}) 
\end{align}
where $r_<$ and $r_>$ correspond to min(r,$r_i$) and max(r,$r_i$) respectively, $H_n$ is the basis function of the biharmonic operator $H_n(x)=\frac{x^{2+n}}{n} \left(\frac{1}{1+x^2}-\frac{1}{1+n } {}_2F_1 \left(2,1+n,2+n,-x^2\right)\right)$ and ${}_2F_1$ is the hypergeometric function with $\frac{1}{1+n} {}_2F_1(2,1+n,2+n,x)=\sum_{i=0}^{\infty}\frac{i+1}{n+i+1} x^i$. $U({\bf x})$ denotes the homogeneous solution of biharmonic($\Delta^2$) operator,
\begin{equation*}\label{U}
U({\bf x})=c_0+d_0\log(1+\tan(\frac{r}{2})^2) + \sum_{n=1}^{\infty}\frac{1}{n}\left(c_n \tan(\frac{r}{2})^n+ d_n H_n(\tan(\frac{r}{2}))\right)\cos(n\phi)
\end{equation*}
with the coefficients
\begin{eqnarray}\label{coefs}
d_0&=&-\log(b)+\frac{1}{2} \left(1+b^{-2}\right) \log\left(\frac{1+b^2}{1+x \rho^2}\right) \nonumber\\
c_0&=&-\text{Li}_2(-b^2)-\log(b)\left(\log(1+b^2)+\log(1+x\rho^2)\right) -d_0 \log(1+b^2) \nonumber\\
d_n&=&\frac{1- \left(1+b^{-2}\right)x \rho^{-n}n H_n(x \rho)}{1-\left(1+b^{-2}\right) b^{-n}nH_n(b)} \left(\frac{x \rho}{b}\right)^nb^{-n} \nonumber\\
c_n&=&\frac{b^{-2n}}{n \big(b^{2+n}-(1+b^2) n H_n(b)\big)} \bigg({x\rho}^n \Big(-b^{2+n}+n H_n(b)
+ b^n n H_n(b^{-1})\left(b^{2+n}-(1+b^2) n H_n(b)\right)\Big)+b^{2+n} n H_n(x\rho)\bigg), \nonumber \\
\end{eqnarray}
with $b=\tan(\frac{\theta_c}{2})$ and $x \rho=\tan(\frac{r_i}{2})$.  

Numerical calculations for ground state configurations are computed using $n=80$ modes in the multipole expansion.  Figure S1(a) shows the convergence of the elastic energy of a single defect near the cap edge with increasing number of modes in the multipole expansion.  Figure S1(b) illustrates that the position of $r_{min}$, corresponding to the optimal position for the defect, vs. number of modes in multipole expansion.  As depicted in the figure, for $n\leq 80$, $r_{min}$ converges to $0.62W$.

\begin{figure}[H]
\centering
\label{modes}
\includegraphics[width=0.9\linewidth]{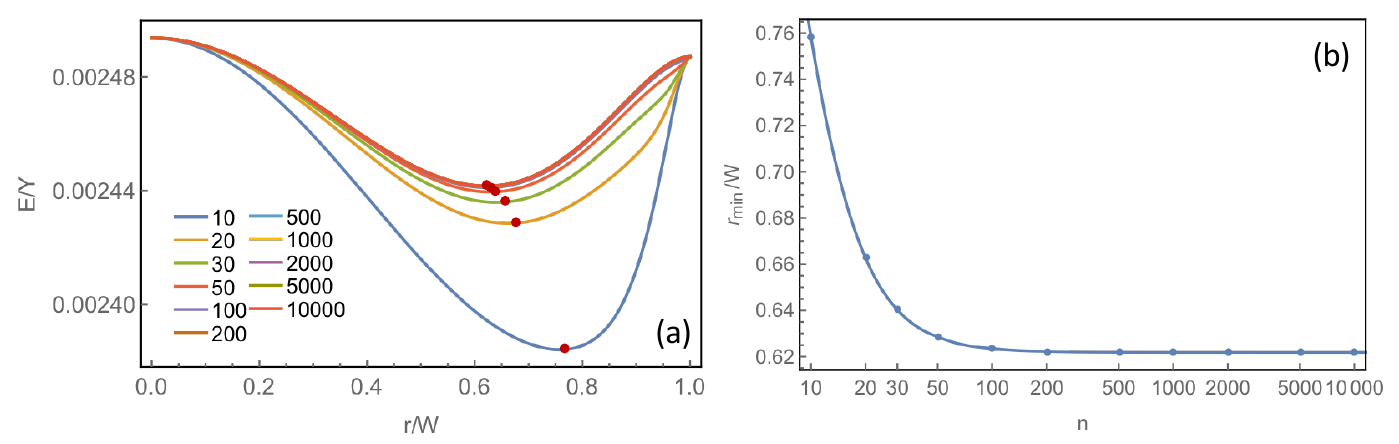}
\caption{The impact of modes in multiple expansion (Eq.~S12) when the defect is near the cap edge.  The cap size is $\theta_c=0.82$.  (a) Elastic energy as a function of single defect position with increasing number of modes (n=10,20..100..10000) (Eq.~S12). (b) The position of minima from (a) as a function of the mode number. The effect number of modes is negligible beyond $n=80$ where the optimal position for the addition of defect remains $r_{min}=0.62W$.}
\end{figure}

\newpage

\subsection{Comparison of defect energetics}

\begin{figure}[H]
\centering
\includegraphics[width=0.6\linewidth]{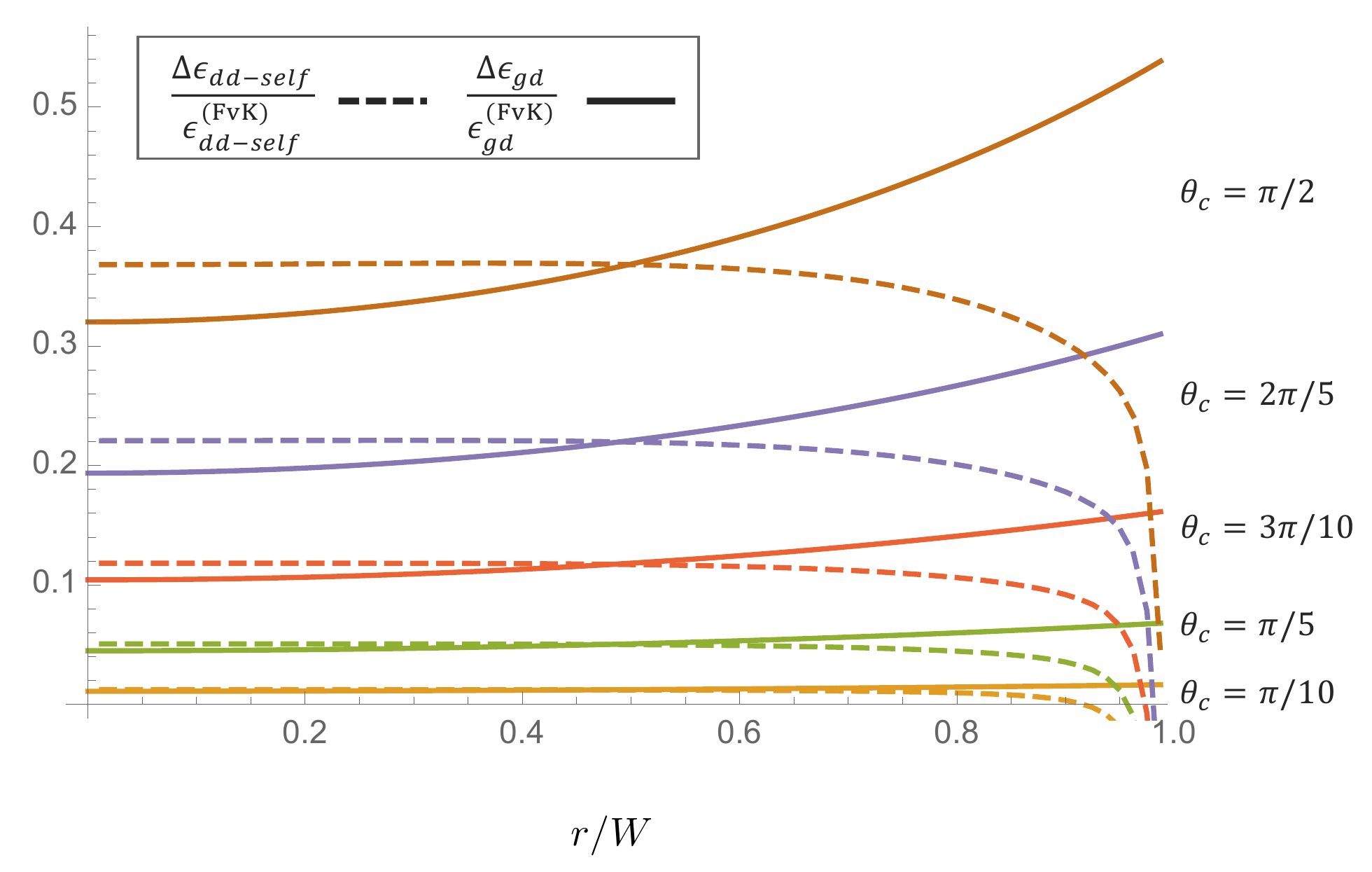}
\caption{Comparison of defect-self energy and defect-curvature terms in FvK vs. LF. The dashed lines correspond to the difference of defect self energy $\Delta \epsilon_{dd-self}/\epsilon^{\rm(FvK)}_{dd-self}$ with $\Delta \epsilon_{dd-self}=
\epsilon^{\rm(LF)}_{dd-self}-\epsilon^{\rm(FvK)}_{dd-self}$, and the solid lines are the difference of defect-curvature interaction $\Delta \epsilon_{gd}/\epsilon^{\rm(FvK)}_{gd}$ with $\Delta \epsilon_{gd}=\epsilon^{\rm(LF)}_{gd}-\epsilon^{\rm(FvK)}_{gd}$.}
\label{compare}
\end{figure}

In Fig.~\ref{compare} we compare the relative differences in the radial dependence of disclination self energy, $\epsilon_{dd-self}$, and defect-curvature interaction, $\epsilon_{dg}$, predicted by FvK and covarient theories.

As summarized in main text, we argue that the enhanced (more negative) defect-curvature attraction at the edge for the LF over FvK derives from higher (and more accurate) geometric compression of the cap edge.  In  the exact spherical embedding (captured by the LF) a hoop of perimeter $2 \pi \rho$ (i.e. at geometric radius $\rho$ from the pole) is compressed to $2\pi R \sin(\rho/R)$ on the sphere.  In comparison, the FvK theory approximated a small slope approximation spherical geometry as a {\it paraboloid} with curvature radius $R$ at its origin. The same hoop at $\rho$ is subjected to a smaller degree of compression when projected onto this small-slope parabolic approximation, and hence, the curvature generated compression is {\it larger} in LF in comparison to the FvK (as is visible in Fig. 4 of ref. \cite{Li2019}) and Fig.~S3(b).  Because the dominant elastic effect of positive disclinations is their tensile far-field stress, we argue that the enhanced geometric compression in the exact spherical embedded underlies the greater near-edge defect-curvature in the LF shown in Fig.~\ref{compare}.

\begin{figure}[H]
\centering
\includegraphics[width=0.6\linewidth]{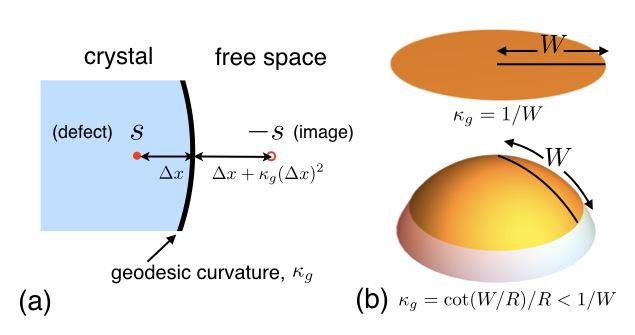}
\caption{In (a), a schematic of the disclination and virtual image charge near to a free boundary of geodesic curvature $\kappa_g$, and (b) and illustration of the ``geodesic flattening" of disc boundaries by spherical geometry.}
\label{geodesicedge}
\end{figure}

The relative {\it depression} of the repulsive self-energy in the covarient theory, relative to FvK, can be understood by considering the limiting form of the self-energy as disclinations approach the edge.  Assuming that the defect distance $\delta x$ from the free boundary is much smaller than either cap size or sphere radius, the dominant contributions of to the self-energy stress are generated by a near-field zone, due to the screening effect of the free boundary.  To a first approximation, the self-stress generated by this defect may be understood by considering the separation between the defect at $r=W-\Delta$ and a virtual defect, of opposite sign, that sits outside of the cap and leads to cancellation of normal stresses on the boundary.  On length scales much smaller than $R$ the cap appears planar, hence we my use the results from FvK theory which show that a disclination (charge $-q_\alpha$) sits at a position $r\simeq W + \delta + \delta^2 \kappa_g$~\cite{grason_pre_12}, where $\kappa_g$ is the geodesic curvature of the edge (see schematic in Fig.~\ref{geodesicedge}).  Hence, the separation between the virtual pair, $2 \Delta x + \delta^2 \kappa_g$, grows with geodesic curvature, leading to a reduction in the far-field screening of defect stress.  The geodesic curvature of spherical caps $\kappa_g=\cot (W/R)/R$ is less than that of a plane of the same radius ($1/W$), and hence, we can expect {\it stronger} boundary in the covariant theory than the FvK screening due to the ``geodesic flattening'' of perimeters.  We argue that this effect, in combination with geometrically the non-linear dependence of the integrated near-field zone within $~\Delta x$ from defect on the spherical geometry, accounts for the near-edge weakening of repulsive self-energy in the covarient elastic theory.

\section{Simulation}

\subsection{Method}
\begin{figure}
\centering
\includegraphics[width=1\linewidth]{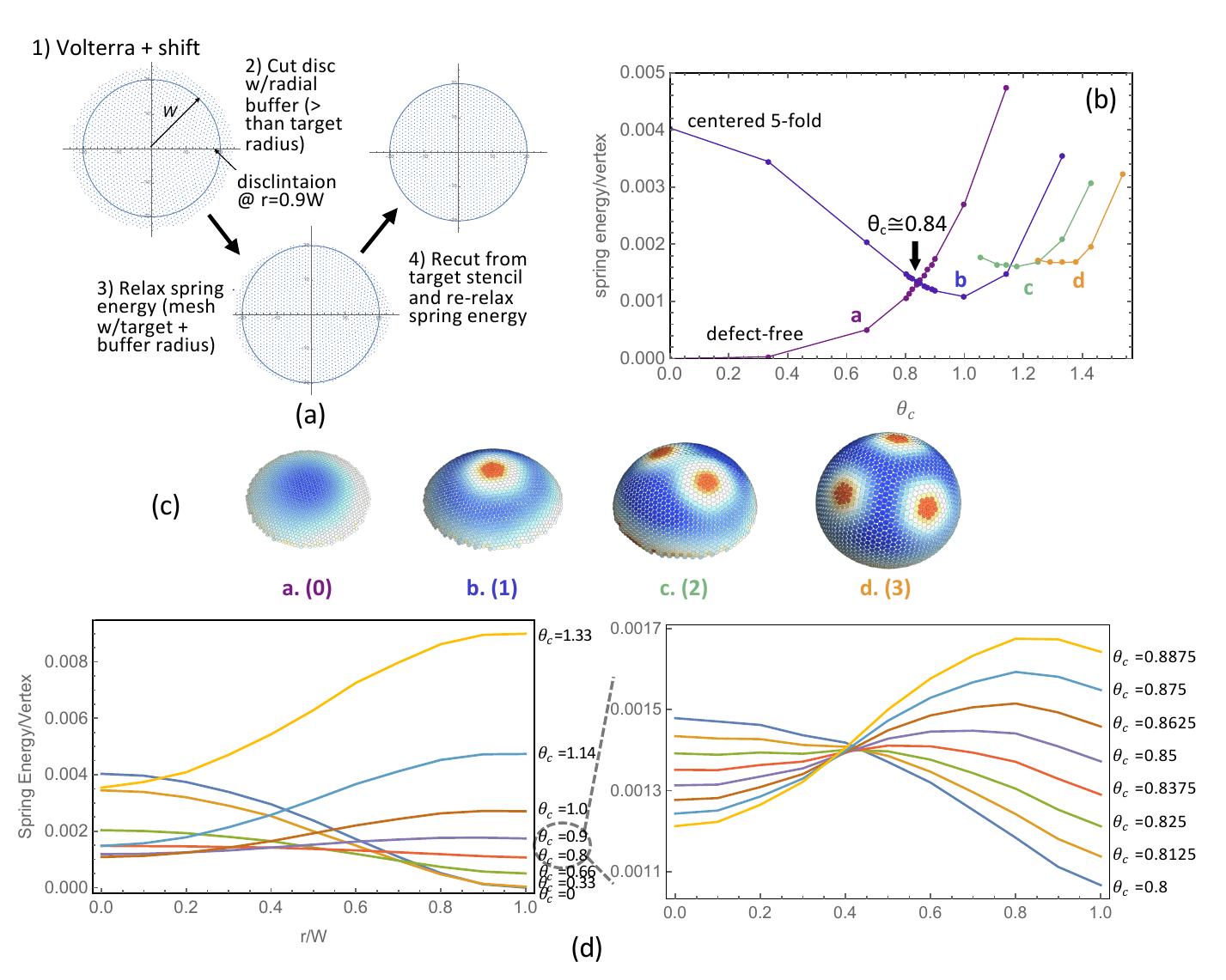}
\caption{Simulation figures.
(a) Volterra operation for mesh generation.
(b) Ground states energy from the simulation, the lines from left to right include zero, 1, 2, 3 disclinations. The defects configurations follow 5-fold, 2-fold, 3-fold icosahedral symmetry. The caps show the strain distribution for each case.
(c) Stress contour plots from the simulation with zero, 1, 2, 3 disclinations.
(d) Left: Spring energy per vertex as a function of single defect location. The cap size ranges from 0 to 1.33, the fine tuned cap sizes between 0.8 and 0.9 are shown on the right side, indicating a boundary-center defect transition around $\theta_c\sim0.84$.
}
\label{sim}
\end{figure}

Our simulation method of the NS model~\cite{nelson_seung}, builds on previous implementations of this model with 5-fold disclinations in a hexagonally-coordinated mesh in controlled locations~\cite{Travesset2003,grason_prl_10}.  In the simulation we first cut a circular disc with one to three disclinations (created via the Volterra operation) each of which corresponds to the removal of $60^\circ$ wedge extending radially outward (created via the Volterra operation).  The 2- and 3-disclination states are constructed by 2-fold and 3-fold arrangements of defects at a fixed radius from the cap center.  Following the cut, a simple azimuthal distortion around each disclination center is imposed to nearly ``close'' each so that retriangulation produces 6-fold coordination at all vertices except the disclination center (5-fold) and the boundary vertices. We then relax the elastic energy (for a planar configuration) via steepest decent.   Depending on the proximity of the defects to the free boundary, this initial relaxation leads to significant distortion away from a circular shape. To mitigate this effect, and its impacts on the predicted elastic energy of circular caps, we first add a ``buffer'' zone and recut the circular disc from the relaxed planar configuration to achieve the target value of $W$.  Upon re-relaxation of the spring energy with the buffer removed, the cap edge retains a nearly circular shape (Fig.~\ref{sim}a).  To consider variable curvature, we project the planar mesh orthographically onto a sphere of fixed radius $R$, and reminimize the spring energy.

\subsection{Results}

The results of the simulation where we choose $W=20 a$ are shown in Fig.~\ref{sim}d.  The transition from defect-free to favorable defect occurs between $\theta_c = 0.8$ and 0.9 (about 0.825) and the favored defect position still appears to be at the center. 

\begin{figure}[H]
\centering
\includegraphics[width=0.9\linewidth]{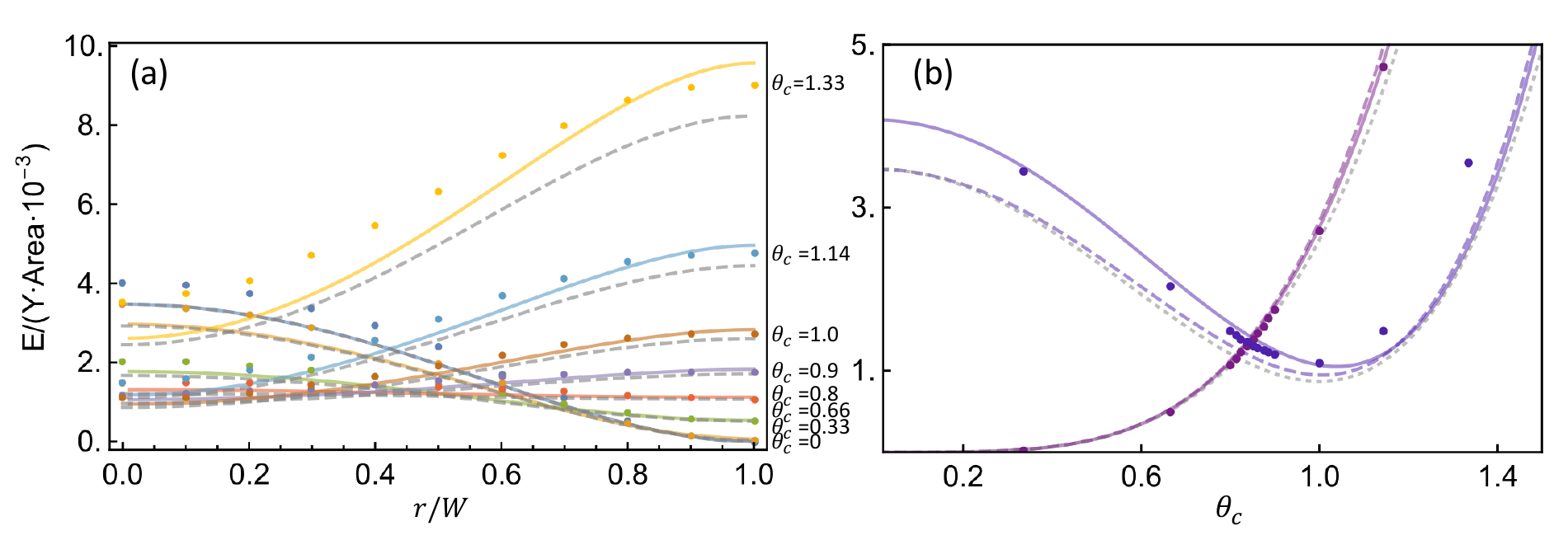}
\caption{
(a) Comparison of single defect energy in simulation (dots), LF (solid lines) and FvK (dashed lines).
(b) Comparison of defect-free and single center-defect energy in simulation (dots) vs. exact solution (solid), LF (dashed) and FvK (dotted).}
\label{singlecomp}
\end{figure}

The comparison to the FvK theory and LF is shown in Fig.~\ref{singlecomp}, which illustrates some considerable ($\sim 10$\%) deviations for when the disclination is near the center. The deviation decreases as the defects approach the boundary for sufficiently small curvatures, which indicates an effect of the non-linear elastic terms and the failure of the small strain approximation.  In particular, both FvK and covariant theories retain only the leading, linear contributions to in-plane strains deriving from the disclinations, neglecting higher-order corrections from large rotations they imposed.  Such errors are reduced if the exact solution is solved as in Ref.~\cite{Li2019}. Nevertheless, many of the quantitative predictions seem to hold (i.e. the curvature threshold and general form of elastic energy vs. defect position) and the exact solution will not be pursued in this article.  Note that the maximum is found in the simulation, which was not significantly be reduced by increasing the mesh points within computational limits of the method.  This discrepancy persists due to combination of two effects:  i) residual anisotropy from cutting a ``circular" sheet from a crystalline lattice; and ii) the incorporation of higher order strain contributions implicit in the NS model but not included in the linear elastic theories.  

Finally, we construct multiple-disclinations, Fig.~\ref{sim}b, shown also in the main text, which matches well with the LF. Both simulation and FvK results show the elastic energy “lags behind” the perfect curvature screening, only LF incorporates several transitions that match the perfect screening.

\subsection{Elastic stress}

To obtain the stress distribution in the shell obtained in simulations, we calculated  the trace of stress tensor (see Fig. 1A), which is
\begin{equation}
\sigma^{ij}_\alpha=\frac{k}{2A}\sum_{\beta}(r_\alpha^i-r_\beta^i)(r_{\alpha\beta}-a)\frac{r_\alpha^j-r_\beta^j}{r_{\alpha\beta}},
\end{equation}
where $k$ is the spring constant, $A=Z \frac{1}{3}\frac{\sqrt{3}}{4}a^2$ is the average area of each vertex with $Z$ the coordination number, and $r_{\alpha\beta}$ is the distance between the neighbors $\alpha$ and $\beta$. Setting $k=\frac{\sqrt{3}}{2} Y$, the trace of stress becomes
\begin{equation}
\sigma^{kk}_\alpha=3Y\frac{1}{Z}\sum_{\beta}(r_{\alpha\beta}-a),
\end{equation}
where $Y$ is the Young Modulus.

\section{Supplementary Figures and Table}
\begin{figure}[H]
    \centering
    \includegraphics[width=0.5\linewidth]{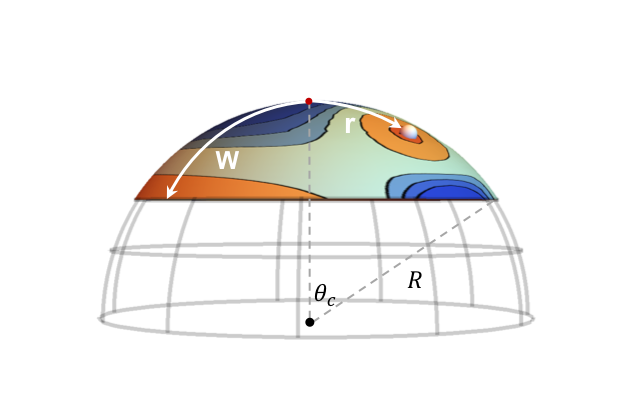}
    \caption{A spherical cap with spherical radius $R$ and geodesic size $W=\theta_c R$.}
\end{figure}
\begin{figure}[H]
\centering
\includegraphics[width=0.6\linewidth]{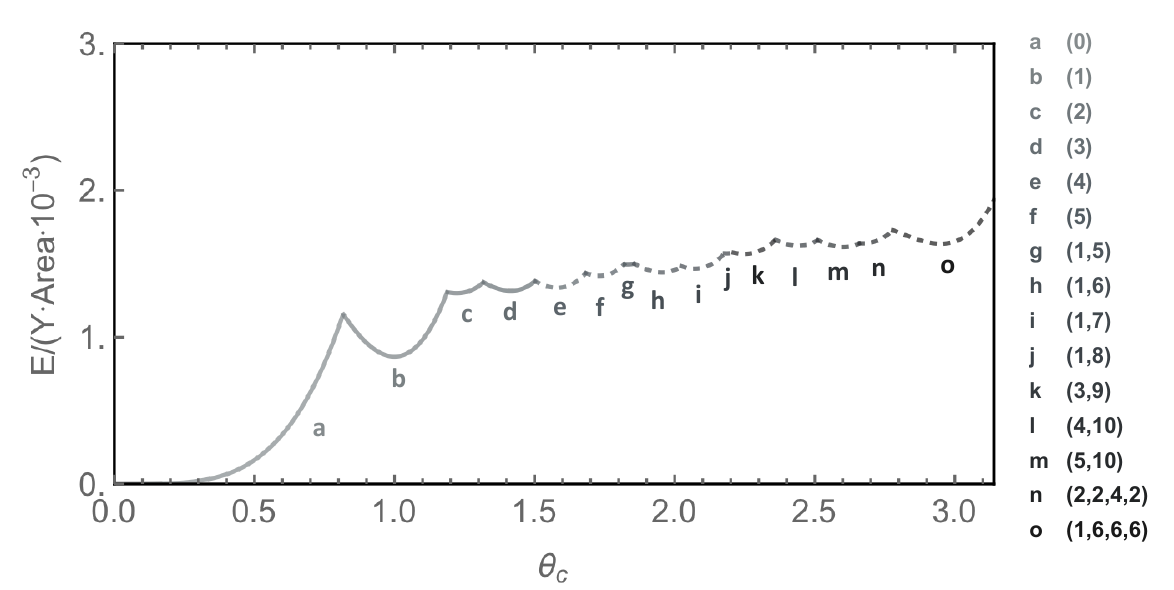}
\caption{The ground states emerged from FvK theory.}
\end{figure}

\begin{figure}[H]
\centering
\includegraphics[width=0.7\linewidth]{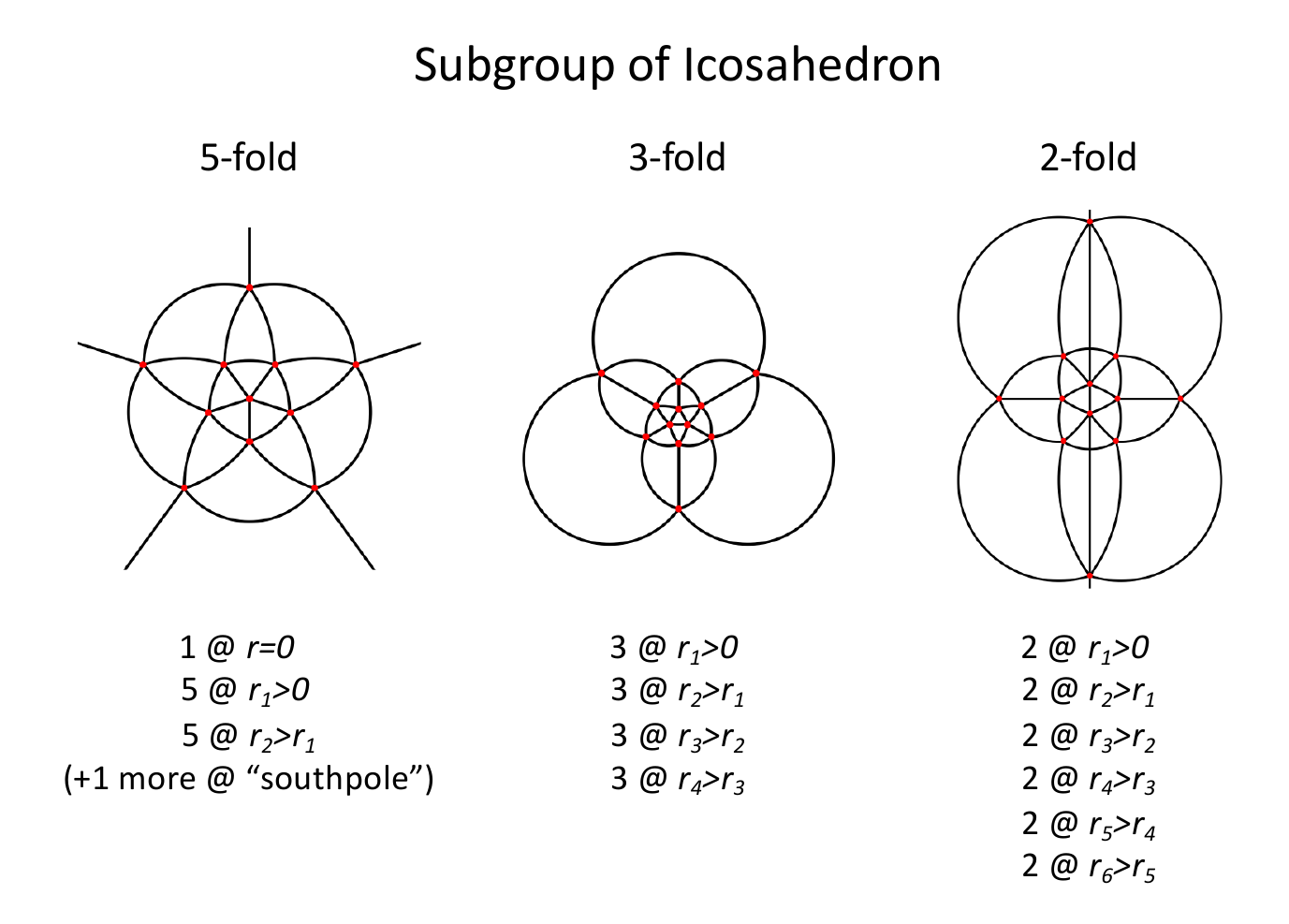}
\caption{The stereographic projection of an icosahedron from the 5-fold, 3-fold and 2-fold symmetry. The disclination configurations of the full sphere in each symmetry group are (1,5,5,1), (3,3,3,3) and (2,2,2,2,2,2).}
\end{figure}

\begin{table}[H]
\includegraphics[width=\linewidth]{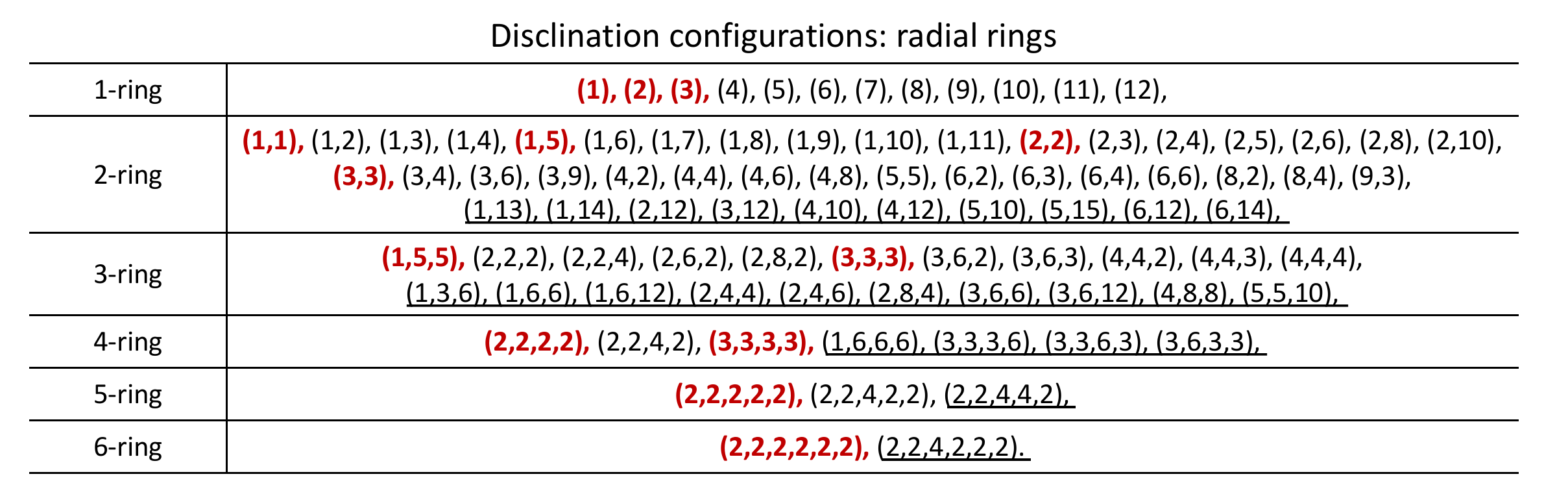}
\caption{Disclination configurations arranged in radial rings. We consider configurations composed of upto 6 nested disclination rings, and defects are evenly spread azimuthally within each ring (i.e. a ring of $p$ defects has $p$-fold rotation symmetry). The icosahedral subgroup defects configurations are indicated in the red color, and the extra configurations used in FvK theory are underlined with total number larger than 12.}
\end{table}

\bibliography{bibfile}